\documentstyle[12pt,epsf]{article}

\newfont{\Bbb}{msbm10 scaled 1200}     
\newcommand{\mathbb}[1]{\mbox{\Bbb #1}}

\newcounter{apps}

\newcounter{prs}[section]

\newcounter{cors}

\newcounter{figs}

\newcounter{th}
\newcommand{\Th}{\par\stepcounter{th}%
{\noindent\bf Theorem \arabic{th}.}\it~}
\newcommand{\Df}{{\noindent\bf Definition. }}

\newcommand{\be}{\begin{equation}}
\newcommand{\ee}{\end{equation}}
\newcommand{\bea}{\begin{eqnarray*}}
\newcommand{\eea}{\end{eqnarray*}}
\newcommand{\beaa}{\begin{eqnarray}}
\newcommand{\eeaa}{\end{eqnarray}}

\newcommand{\ba}{\begin{array}}
\newcommand{\ea}{\end{array}}

\newcommand{\lb}{\label}
\newcommand{\g}{\gamma}
\newcommand{\G}{\Gamma}
\newcommand{\ra}{\rightarrow}

\newcommand{\wt}{\widetilde}
\newcommand{\td}{\tilde}
\newcommand{\e}{\epsilon}
\newcommand{\al}{\alpha}
\newcommand{\bt}{\beta}
\newcommand{\p}{\partial}

\newcommand{\vp}{\varphi}

\newcommand{\HH}{{\mathcal{H}}}
\newcommand{\Hd}{{\mathcal{H}}^*}

\begin{document}

\begin{flushright}
ITEP-TH-43/03\\
\end{flushright}

\medskip

\begin{center}
\bigskip
{\large\bf Leading RG logs in $\vp^4$ theory}

\bigskip
\bigskip

Dmitry Malyshev%
\footnote{
malyshev@gate.itep.ru}

\bigskip

{\small\it
Institute of Theoretical and Experimental Physics\\
and Moscow State University, Moscow, Russia\\
}

\bigskip

\bigskip
{July 2003}

\bigskip
\bigskip

\end{center}

\begin{abstract}
We find the leading RG logs in $\vp^4$ theory
for any Feynman diagram with 4
external edges.
We obtain the result in two ways.
The first way is to calculate the relevant terms in Feynman integrals.
The second way is to use the RG invariance based on
the Lie algebra of graphs introduced by Connes and Kreimer.
The non-RG logs, such as $(\ln s/t)^n$, are discussed.
\end{abstract}

\section{Introduction}

Radiative corrections give the most precise checks of the quantum
field theory.
Calculations of the corrections are based on multiloop
calculations.
Thus a special attention is paid to the multiloop calculations
during the whole history of QFT.
But all of the results are rather involved \cite{Tkachev,Grozin}.
It is a new branch of mathematics
with its own methods and special functions.
We present here a set of results which have a spirit of
multiloop calculations but on
the other hand we need only elementary mathematics and the answers have a
simple form.

The leading logs may be used as a toy model in developing methods
for the calculation of Feynman integrals.
One can calculate the integral and then check the results by the RG
invariance.
Indeed, it is known that the leading logs are determined by
the one-loop
RG invariance \cite{Bog,Itsyk,Col}
(for an elementary discussion of this fact see \cite{Delam}).

The problem is that if we start from a symmetric point, then the
one-loop RG fixes the leading logs
in the symmetric points $s=t=u=\mu^2$,
but the structure of leading logs for
arbitrary external momenta is not fixed.
Furthermore the traditional RG
gives the coefficients only for the sum of
the leading logs of a given power.

The main result of the paper is an explicit calculation of the leading RG logs for
arbitrary Feynman diagrams with 4 external edges.
The result contains two parts:
\begin{enumerate}
  \item We prove that the leading RG logarithm for any Feynman
  diagram depends only on one Mandelstam variable.
  Namely, the leading RG logarithm for an n-loop diagram $\g_n$ has the
  form
\be
F(\g_n)=c(\g_n)\left(\ln\frac{\mu^2}{s}\right)^n,
\ee
  or the same expression with $t$ or $u$, depending on the
  orientation of the diagram.
  \item We find the coefficients
\be\lb{coef1}
c(\g_n)=\sum_{T_{\g_n}}\frac{1}{n_{\g_1}}\cdots\frac{1}{n_{\g_k}},
\ee
where the sum is over maximal rooted trees $T_{\g_n}$
of divergent subgraphs $\g_i\subset\g_n$
and $n_{\g_i}$ is the number of loops in the
graph $\g_i$.
\end{enumerate}
In section 5
we obtain the same result for $c(\g)$ using
a version of RG equations based on
the Connes and Kreimer Lie algebra of graphs \cite{CK1,CK2}.
We find the recursive formula
\be\lb{coef2}
c(\g_{n+1})
=\frac{1}{n+1}\;\sum_{\g_n=\g_{n+1}/\g_1}\;
c(\g_{n}),
\ee
where $\g_{n+1}$ is an $(n+1)$-loop graph,
$\g_1$ is a one-loop subgraph of $\g_{n+1}$,
$\g_n=\g_{n+1}/\g_1$ is the $n$-loop graph,
obtained by the contraction of the one-loop subgraph
$\g_1\subset\g_{n+1}$ into a vertex.
The sum is over all possible $n$-loop graphs $\g_{n}$,
which may be obtained by the contraction of the one-loop subgraphs
in $\g_{n+1}$.

We check that formulas (\ref{coef1}) and (\ref{coef2}) coincide in
several nontrivial examples.

\section{The model}

We consider the massless $\vp^4$ theory in $d$-dimensional
euclidian space-time with the lagrangian
$$
L=\frac{1}{2}(\p_\mu\vp)^2+\frac{16\pi^2}{4!}\vp^4.
$$
For the regularization of UV divergences we use the
dimensional regularization with $d=4-2\e$ \cite{'tHdimreg}.

In massless theory the one-loop correction to the two-point
function vanishes.
This fact has two consequences:
\begin{enumerate}
  \item We may ignore the $Z$-factors in front of the four-point
  vertex function, since they will play a role only in two-loop
  $\bt$-function, thus giving a contribution to the next-to-leading logs.
  \item Diagrams with self energy insertions do not give leading RG logs.
\end{enumerate}
Further we consider only 1PI graphs with four external edges
without self energy insertions.
We assume non-zero external momenta, thus our
diagrams have no infrared divergences \cite{Itsyk}.

The one-loop diagram has the form

\begin{figure}[h]
\begin{center}
\leavevmode
\epsfxsize 100pt
\epsffile{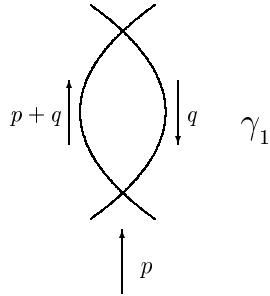}
\caption{\bf One-loop diagram}
\end{center}
\end{figure}

The corresponding Feynman integral is
\be
F_{\rm I}=\int\frac{d^dq}{(q+p)^2 q^2}\;.
\ee
Here $p^2$ is one of the three Mandelstam variables $s,t$ or $u$.
In parametric representation after the integration over $q$ we get
\be
\lb{1}
F_{\rm I}=\int_{0}^{1}dx\int_{0}^{\infty}\frac{d\al}{\al}
    (\mu^2\al)^\e e^{-\al x(1-x)p^2}.
\ee
The integration over $\al$ gives
\be
\lb{1l}
F_{\rm I}=\G(\e)\left(\frac{\mu^2}{p^2}\right)^\e+\ldots
\ee
The dots denote the terms which are not important for
the leading RG logs.

\section{Two-loop diagram}

The problem in multiloop calculations is the existence of
divergent sub-integrals.
In order to get a finite answer one has to subtract the
sub-divergences and then the overall divergence.
The structure of the subtraction procedure is the subject of
Bogolubov's R-operation \cite{Bog}.

The main property of divergent terms is the locality in
$x$-space.
The locality means that the corresponding diagram (sub-diagram) is
reduced to a point, i.e. to a vertex.
Thus the divergence may be subtracted by a local counterterm in
the vertex of a smaller diagram.

In parametric representation the integrals are divergent near the zero
of the $\al$'s corresponding to the divergent sub-diagram
\cite{Bog,Itsyk,Col,Zav}.
Consequently the divergent terms are local in $\al$-space.
It will be shown that the terms contributing to the leading logs
go together with divergent terms and have the local structure in
$\al$-space.
Extraction of the leading logarithm from a sub-diagram reduces the
sub-diagram to a vertex.
The leading logarithm has the form $(\ln\mu^2\al)^l$, where $l$ is
the number of loops and $\al$ is the overall scaling for the
sub-diagram.
Thus from the point of view of the leading logs the sub-diagram
has the form of a vertex with the interaction $(\ln\mu^2\al)^l$.

In the calculation we
decompose the singular integrals into  singular and
nonsingular parts.
The leading RG logs are extracted from the most singular part,
which contains all the sub-divergences of the Feynman integral.
The products of singular and nonsingular parts will give
sub-leading RG logs or non-RG logs.
Let us stress that we will study only the leading RG logs, the
products of RG and non-RG logs,
such as  $(\ln \mu^2/t)^m(\ln s/t)^k$,
are not the leading RG logs.

The decomposition of a singular integral on a divergent local part
and a convergent part is connected with the regularization of
singular distributions \cite{Gelfand}.
This analogy was used to express the R-operation in terms of
distributions \cite{Tkachev}.
We also explore this analogy
in order to subtract the divergences,
but we work in $\al$-space rather then in $x$ or $p$ spaces.

The nontrivial two-loop diagram has the form shown in figure 2.
\begin{figure}[h]
\begin{center}
\leavevmode
\epsfxsize 100pt
\epsffile{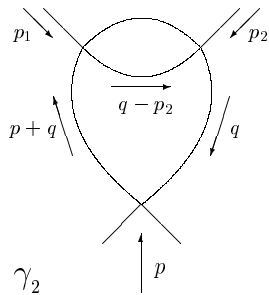}
\label{2loop}
\caption{\bf Two-loop diagram in $s$ channel}
\end{center}
\end{figure}
The momentum $q-p_2$ is the external momentum for
the one-loop sub-diagram.
We don't put the internal momentum for the one-loop sub-diagram in the
figure.
The corresponding Feynman integral is
\be\lb{2l}
F=\int\frac{d^dq}{(q+p)^2 q^2}
\G(\e)\left(\frac{\mu^2}{(q-p_2)^2}\right)^\e+\ldots,
\ee
here we denote $p=-(p_1+p_2)$ and
use the answer (\ref{1l}) for the one-loop sub-diagram.
In parametric representation
$$
F=\int_0^\infty d\al_1d\al_2\frac{d\al_3}{\al_3}(\mu^2\al_3)^\e
\int d^dq e^{-\al_1(q+p)^2-\al_2q^2-\al_3(q-p_2)^2}.
$$
We integrate over $q$ and choose the new variables
\be
\ba{l}
\al_1=\al x,\\
\al_2=\al y,\\
\al_3=\al z,\\
\al\in(0,\infty),\;x,y,z\in(0,1)\\
x+y+z=1.
\ea
\ee
Then
\be\lb{2loop}
F=
\int_{0}^{\infty}\frac{d\al}{\al}(\mu^2\al)^\e
\int_{0}^{1}dx\int_{0}^{1-x}\frac{dz}{z}(\mu^2\al z)^\e
 e^{-\al Q},
\ee
where
$$
Q=xyp^2+xzp_1^2+yzp_2^2,\;\;\;y=1-x-z.
$$

The integral has two singularities, $\al=0$ -- the
overall divergence
and $z=0$ -- the sub-divergence.
We decompose the integral over $z$ into a singular part and
non-singular parts.
In general case we have the following integral
\be\lb{I}
I=\int_0^a\frac{dz}{z}z^\e f(z),
\ee
where $f(z)$ is a regular function of $z$ .
Integrating by parts and expanding in $\e$ we get
\be\lb{dec}
I=
\frac{1}{\e}f(0)+\ln(a)f(a)
-\int_0^a\ln(z)\frac{df}{dz}dz+{\rm O}(\e).
\ee
Note, that the singular in $\e$ term is local in $z$,
where $z$ is the parameter corresponding to the divergent
sub-diagram.
We are going to show, that the first term gives the leading
logarithm and together with
the second term may contribute to the beta-function,
the last term does not depend on the renormalization parameter
$\mu$, thus it may give only a non-RG logarithm.

For convenience, we introduce a new parameter $\e_1$ for the
sub-divergent integral,
then the first term in (\ref{dec}),
substituted in (\ref{2loop}), has the form
\be\lb{}
F^{(1)}=
\int_{0}^{\infty}\frac{d\al}{\al}(\mu^2\al)^\e
\frac{1}{\e_1}\int_{0}^{1}dx(\mu^2\al)^{\e_1}
 e^{-\al x(1-x)p^2},
\ee
here we have used
$$
\left.Q\right|_{z=0}=x(1-x)p^2.
$$
The dependence on $p_1^2$ and $p_2^2$ vanishes because of the
locality in $z$.
The counterterm for the sub-divergence in (\ref{2l}) has the form
\be
C_{\rm I}
=-\frac{1}{\e_1}
\int\frac{d^dq}{(q+p)^2 q^2}.
\ee
In parametric representation
\be
C_{\rm I}=
-\frac{1}{\e_1}
\int_{0}^{\infty}\frac{d\al}{\al}(\mu^2\al)^\e
\int_{0}^{1}dx  e^{-\al x(1-x)p^2}.
\ee
After the subtraction we get
\be\lb{}
\wt{F}^{(1)}
=\int_{0}^{\infty}\frac{d\al}{\al}(\mu^2\al)^\e
\frac{1}{\e_1}((\mu^2\al)^{\e_1}-1)\int_{0}^{1}dx
 e^{-\al x(1-x)p^2}.
\ee
It may be shown that the finite answer does not depend on the way
we take the limit $\e,\e_1\ra 0$.
It is convenient to take the limit $\e_1\ra 0$ first, then
\be
\wt{F}^{(1)}
=\int_{0}^{\infty}\frac{d\al}{\al}(\mu^2\al)^\e
\ln(\mu^2\al)\int_{0}^{1}dx
 e^{-\al x(1-x)p^2}.
\ee
We see that the integral looks as the one-loop integral (\ref{1}) with
$\ln(\mu^2\al)$ inserted in the upper vertex, where we have had
the divergent sub-diagram.

Integration over $\al$ gives
\be
\wt{F}^{(1)}
=\p_\e\left[\G(\e)\left(\frac{\mu^2}{p^2}\right)^\e\right]
+\ldots
\ee
Expanding in $\e$ we find the leading RG logarithm
\be\lb{tl}
\wt{F}^{(1)}
=-\frac{1}{\e^2}+\frac{1}{2}\left(\ln\frac{\mu^2}{p^2}\right)^2+\ldots
\ee
For the second term in (\ref{dec}) we note that $a=1-x$ and
$$
\left.Q\right|_{z=1-x}=x(1-x)p_1^2,
$$
then after the integration over $\al$
\bea
F^{(2)}
=-\G(\e)\left(\frac{\mu^2}{p_1^2}\right)^\e+\ldots
\eea
This term may give a contribution only to the beta-function.

The last term of (\ref{dec}) gives a regular expression in (\ref{2loop}),
since
$$
\frac{d}{dz}e^{-\al Q}\sim\al
$$
and the singularity at $\al=0$ vanishes
\be
F^{(3)}
=\int_{0}^{\infty}d\al \int_{0}^{1}dx  \int_{0}^{1-x}dz\ln z
\frac{dQ}{dz}e^{-\al Q}.
\ee

If we consider the diagram in $t$ channel and take the limit
\be\lb{lim}
p_1^2\sim p_3^2\sim s\sim u>>t=(p_1+p_3)^2,
\ee
then the third term will give
the non-RG logarithm
\be
F^{(3)}_2
=-\frac{1}{2}\left(\ln\frac{s}{t}\right)^2+\ldots
\ee
The first two terms in (\ref{dec}) are 'local' in $z$, thus
their structure does not depend on the asymptotic of the external
momenta.
The last term is 'non-local' in $z$ and it has different behavior
in different asymptotics.
In general case of a multiloop diagram the non-RG logs will be given
by combinations of 'second' and 'third' terms for sub-divergences.
Some estimates of the non-RG logs may be found in
\cite{Weinberg,Ginzburg}.

We see that the decomposition efficiently divides the integral
into RG and non-RG parts.
Further we will not consider the second and the third terms, since
they do not contribute to the leading RG logs.

\section{Calculation of the leading RG logs}

In this section
we formulate the main theorem, i.e. we prove that
the leading RG logarithm depends on one Mandelstam variable
and find the coefficients.

\Df A diagram is called reducible if the Feynman integral splits into a product
of Feynman integrals for the sub-diagrams.

{\Th
Let $\g_n$ be an irreducible n-loop diagram.
The most singular term in the Feynman integral
after the subtraction of
sub-divergences has the form}
\be\lb{anz}
F(\g_n)\sim \p^{n-1}_\e
\left[\G(\e)\left(\frac{\mu^2}{p^2}\right)^\e\right]
\ee

We have seen that the assertion is true in the case of $n=1,2$.
Assume that it is true for all $i\leq n$ and prove it for $i=n+1$.

Any diagram $\g_{n+1}$ with $n+1$ loops has the form
shown in figure 3.
\begin{figure}\label{n1loop}
\begin{center}
\leavevmode
\epsfxsize 100pt
\epsffile{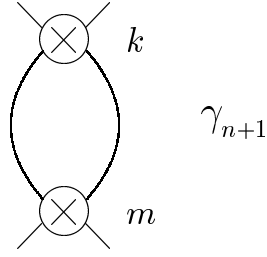}
\caption{\bf Form of the diagram with $n+1=k+m+1$ loops}
\end{center}
\end{figure}
Here the circles represent $k$-loop and $m$-loop sub-diagrams.

Since the number of loops in $\g$ is $m+k+1=n+1$,
then $k,m\leq n$.
Assume that the sub-diagrams are irreducible, then we can use
expression (\ref{anz}) for the sub-integrals.
In this case the most singular term is
\be\lb{msing}
F=\int\!\!\frac{d^dq}{(q+p)^2\:q^2}
\;\p^{m-1}_{\e_1}\!\!\left[\G(\e_1)\!\!\left(\frac{\mu^2}{(q-p_3)^2}\right)^{\e_1}\right]
\;\p^{k-1}_{\e_2}\!\!\left[\G(\e_2)\!\!\left(\frac{\mu^2}{(q+p_1)^2}\right)^{\e_2}\right],
\ee
where $p=p_1+p_2$.

In parametric representation the integrals concerning the
sub-divergences have the form (compare with (\ref{I}))
\be\lb{}
I=\p^{m-1}_{\e_1}\p^{k-1}_{\e_2}
\int_0\frac{dz}{z}z^{\e_1}(\mu^2\al)^{\e_1}
\int_0\frac{du}{u}u^{\e_2}(\mu^2\al)^{\e_2}
f(z,u)
\ee
with some regular function $f(z,u)$.

The most singular term in the integral is
\be\lb{sing}
I_{sing}=
\p^{m-1}_{\e_1}\left[\frac{1}{\e_1}(\mu^2\al)^{\e_1}\right]
\p^{k-1}_{\e_2}\left[\frac{1}{\e_2}(\mu^2\al)^{\e_2}\right]
f(0,0).
\ee
If we expand in $\e_1$ and $\e_2$ and subtract the sub-divergences in
(\ref{msing}),
then only the finite term in (\ref{sing}) survives
$$
I_{fin}=
\frac{1}{m}(\ln\mu^2\al)^m
\frac{1}{k}(\ln\mu^2\al)^k
f(0,0),
$$
where
$$
f(0,0)=e^{-\al x(1-x)p^2}.
$$
The whole integral is
$$
F\sim
\int_{0}^{\infty}\frac{d\al}{\al}(\mu^2\al)^\e
\frac{1}{m}\frac{1}{k}\ln(\mu^2\al)^{m+k}
e^{-\al p^2},
$$
where we ignore the parameter $x$, since it is not important
for the leading logarithms or singularities.
Note, that the integral again has the form of the one-loop integral with
$1/k(\ln\mu^2\al)^k$ in the upper vertex and
$1/m(\ln\mu^2\al)^m$ in the lower vertex.

In the case of reducible sub-diagrams only the numerical
coefficient changes,
it will be the product of $1/n_l$, where $n_l$ are the numbers of
loops in the irreducible parts of the sub-diagram.

Integrating over $\al$ we find
the most singular term and the leading logarithm
\bea
F&\sim&
\frac{1}{m}\;\frac{1}{k}\;
\p^{n}_\e
\left[\G(\e)\left(\frac{\mu^2}{p^2}\right)^\e\right],\\
F_{ll}&\sim&
\frac{1}{n}\;\frac{1}{m}\;\frac{1}{k}
\left(\ln\frac{\mu^2}{p^2}\right)^n.
\eea

Note, that the coefficient for the leading logarithm
is proportional to one over the number
of loops in the diagram and in the two sub-diagrams.
If the sub-diagrams have divergent
sub-sub-diagrams, then the coefficient is
also divided by the number of loops in these sub-sub-diagrams.
Thus the coefficient in front of the leading RG logarithm should
be proportional to the product of $1/n_l$, where $n_l$ are the
numbers of loops in some sub-diagrams of initial graph.
For example, if an $n$-loop diagram $\g$ has only one $(n-1)$-loop
sub-diagram, this sub-diagram has only one $(n-2)$-loop
sub-sub-diagram etc., then
\be\lb{1ch}
c(\g)=\frac{1}{n!}.
\ee
Sometimes a graph may be decomposed into subgraphs in different
ways with different numbers of loops in the subgraphs.
Then the coefficient is a sum over the decompositions.
This situation arises in the case of overlapping divergences,
when there are different intersecting domains of
singularities in the space of parameters $\al$ and one should take
the sum over the domains.

Intuitively the picture is the following.
Consider the graph $\g_3$ in figure 4.

\begin{figure}[h]
\begin{center}
\leavevmode
\epsfxsize 100pt
\epsffile{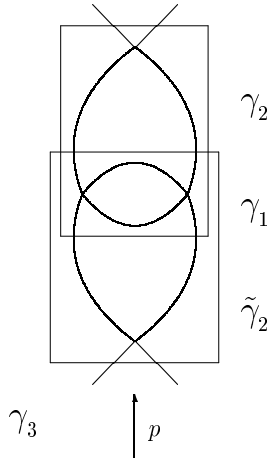}
\label{3loop}
\caption{\bf Three-loop diagram $\g_{_3}$}
\end{center}
\end{figure}

It has two two-loop subgraphs, $\g_2$ -- inside the upper box
and $\td{\g}_2$ -- inside the lower box, which intersect by a one-loop
subgraph $\g_1$ -- inside the intersection of the boxes.

Let $\al$ and $\bt$ be the parameters corresponding to the
subgraphs $\g_2$ and $\td{\g}_2$ respectively.
In the singularity $\al\ra 0$ ($\bt\ra 0$) the upper (lower)
subgraph reduces to a point.
In the picture it looks as if the corresponding box shrinks to a point
\footnote{
The operation of extraction of a subgraph which is inside a box is essential
in the definition of the Hopf algebra of graphs \cite{CK1,GMS}.}.
Since the divergences are overlapping, one should divide the
integration domain into two regions $\al<\bt$ and $\bt<\al$,
the singularity $\al\ra 0$ is in the first region,
while $\bt\ra 0$ is in the second.
From the point of view of the above theorem,
the first case corresponds to the chain of subgraphs
$\g_1\subset\g_2\subset\g_3$,
in the second case $\g_1\subset\td{\g}_2\subset\g_3$.
In each case the coefficient is $1/3!$, thus the final answer,
which is the sum over the domains (chains)
\be\lb{3l}
c(\g_3)=\frac{1}{3!}+\frac{1}{3!}=\frac{1}{3}.
\ee
In general case of overlapping divergences
there is a sum over the maximal rooted trees of
divergent subgraphs \cite{Kr:overlap}.
The trees of divergences were introduced by Zimmermann \cite{Zim:forest}
(in his work the tree is called the forest),
the rooted trees were used by Connes and Kreimer
in the yearly works on the Hopf algebra and
R-operation \cite{Kr:overlap,Connes:tree}.
In our case the trees are maximal, since we are interested only in
the leading logs.

\Df A rooted tree $T_\g$ is corresponding to the graph $\g$,
if every node $A\in T_\g$ corresponds to a subgraph
$\g_A\subset\g$, the root of $T_\g$ corresponds to the graph $\g$,
and for any subgraphs $\g_{A_1}\subset\g_{A_2}$ the corresponding
node $A_2$ lies on the shortest path from $A_1$ to the root
(i.e. for larger subgraphs the corresponding vertices are
closer to the root).

We will consider only divergent subgraphs,
i.e. the subgraphs with four external edges.

\Df A rooted tree $T_\g$ is called maximal if the corresponding
set $\{\g_{A_1}\ldots\g_{A_k}\}$ of subgraphs is maximal,
i.e. for any $\td{\g}\subset\g$ the set
$\{\td{\g},\g_{A_1}\ldots\g_{A_k}\}$ does not correspond to any
rooted tree (the set $\{\td{\g},\g_{A_1}\ldots\g_{A_k}\}$
will correspond to a graph with a loop, which is not a tree).

Note, that every node of a tree has a corresponding subgraph,
but some subgraph may have no corresponding nodes on the tree,
even in the case of the maximal tree.

The result of the above discussion may be expressed as
{\Th
Let $\g$ be a 1PI graph with four external edges and without
self energy insertions.
The coefficient in front of the
leading RG logarithm is
\be\lb{c1}
c(\g)=\sum_{T_\g}\frac{1}{n_{\g_1}}\cdots\frac{1}{n_{\g_k}},
\ee
where the sum is over the maximal trees $T_\g$ of subgraphs in
$\g$,
$\{\g_{1}\ldots\g_{k}\}$ is the set of subgraphs of $\g$
corresponding to $T_\g$ (including the graph $\g$),
and $n_{\g_i}$ is the number of loops in $\g_i$.
}

\section{Lie algebra of graphs}

In this section we consider
the notion of linear space of graphs introduced by Connes and
Kreimer \cite{CK1}.
We need one property of the notion -- the product of graphs (dual
to the coproduct in the Hopf algebra $\HH$ of graphs).
The main result of the section is the derivation of
recursive formula (\ref{coef1})
from the RG equation in the space $\Hd$ \cite{CK2}.

In fact the idea of RG invariance in the space $\Hd$  is very simple.
In different language it is discussed be Peskin and Schroeder
\cite{Peskin} in \S 10.5,
where the divergent diagrams in $s$ channel are divided into three groups and
the divergences cancel separately within each group.
The same arguments give independent RG equations in these three
groups.

The linear space $\Hd$ is a linear space where the basis vectors
are labeled by Feynman diagrams, i.e. every graph is associated
with a basis vector,
the graph and the corresponding vector will be
denoted by the same letter $\g$.

Vertex functions and the beta-function are some vectors in $\Hd$
\cite{CK2}.
For example,
\bea
\G&=&g\g_0-\frac{1}{2}g^2
            \left(
              \ln\frac{\mu^2}{s}\g_1^{(s)}
              +\ln\frac{\mu^2}{t}\g_1^{(t)}
              +\ln\frac{\mu^2}{u}\g_1^{(u)}
            \right)
          +\ldots\\
\bt&=&\frac{1}{2}g^2
            \left(\g_1^{(s)}+\g_1^{(t)}+\g_1^{(u)}
            \right)
          +\ldots
\eea
where\\
$\g_0$ is a vertex, graph with zero loops,\\
$\g_1^{(s)},\;\g_1^{(t)},\;\g_1^{(u)}$
are the one-loop graphs with different orientations.
For our purposes it is convenient to distinguish the orientation
of graphs.

The RG equation for the running coupling is
\bea
\frac{d\bar{g}}{d\ln\mu^2}
&=&\bt(\bar{g})\\
&=&\frac{1}{2}\bar{g}^2
\left(\g_1^{(s)}+\g_1^{(t)}+\g_1^{(u)}\right)+\ldots
\eea

The RG equation for the vertex function is
\be\lb{rg}
\frac{d}{d\ln\mu^2}\G(s,t,u;\bar{g}(\mu),\mu)=0.
\ee
The crucial difference between the usual RG equation and the RG equation in
the space $\Hd$ is that in the first case there is only one equation at a
given power of the coupling while in the second case there is a system of
equations since the vertex function is a vector.

After the substitution of the beta-function in (\ref{rg}) the RG equation
reads
\beaa
\nonumber
\frac{d}{d\ln\mu^2}\G
&=&\frac{1}{2}\bar{g}^2\left(\g_1^{(s)}+\ldots\right)\ast\g_0
-\frac{1}{2}\bar{g}^2\left(\g_1^{(s)}+\ldots\right)\\
\lb{last}
&+&\bar{g}^3\left(\g_1^{(s)}+\ldots\right)
\ast\left(\ln\frac{\mu^2}{s}\g_1^{(s)}+\ldots\right)+\ldots
\eeaa
The dots in the parentheses denote the graphs of $t$ and $u$
orientation.
In order to satisfy equation (\ref{rg}) we have to define the
multiplication of graphs such that
$\g_1\ast\g_0=\g_1$ is a one-loop graph,
$\g_1\ast\g_1=\g_2$ is a two-loop graph etc.
The product of graphs was defined by Connes and Kreimer
\cite{CK1,CK2}.
It looks as an insertion of the graph
in the vertices of the other graph%
\footnote{In fact the product of graphs $\g_1$ and $\g_2$ contains
also the disconnected union $\g_1\sqcup\g_2$.
In order to avoid such terms one should make a projection on
connected graphs when it is needed.}.

The insertion may be expressed in the following way.
Let us draw the graphs together with the couplings in the vertices.

We decompose
\be
\frac{d}{d\ln\mu^2}=\frac{\p}{\p\ln\mu^2}+\bt(g)\frac{\p}{\p g},
\ee
where the partial differentiation in $\mu$ acts only on logs and the
differentiation in $g$ acts on the couplings standing in the vertices.
The operator $\bt(g)\p_g$ substitutes $\bt(g)$ for $g$ in
the vertex.
Since $\bt(g)$ is a vector in $\Hd$, i.e. it depends on graphs,
$\bt(g)\p_g$ substitutes the graphs of $\bt(g)$ in place of the
vertex on which $\p_g$ acts.

\begin{figure}[h]
\begin{center}
\leavevmode
\epsfxsize 300pt
\epsffile{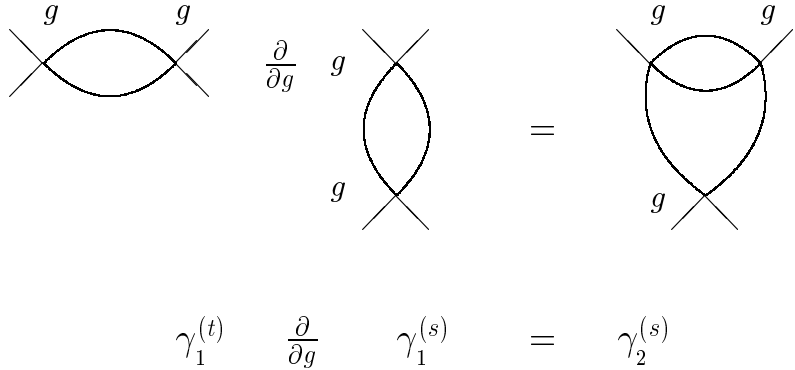}
\label{2l_ins}
\caption{\bf Insertion of  $\g_{_1}^{(t)}$ into  $\g_{_1}^{(s)}$}
\end{center}
\end{figure}

The graph $\g_2^{(s)}$ in the figure may be obtained only by
insertion of $\g_1^{(t)}$ or $\g_1^{(u)}$ in the upper vertex of
$\g_1^{(s)}$.

Now we are going to use the RG equation
\be
\frac{d}{d\ln\mu^2}\G
=\left(\frac{\p}{\p\ln\mu^2}+\bt(g)\frac{\p}{\p g}\right)\G
=0
\ee
and find the leading RG logarithm for $\g_2^{(s)}$.

The part concerning $\g_2^{(s)}$ is
\be\lb{bet}
(\bt_t+\bt_u)\left(\frac{\p}{\p g}g\right)g
\g_1^{(s)}\frac{1}{2}\ln\frac{\mu^2}{s}
=\frac{c(\g_2)}{2}g^3\g_2^{(s)}
\frac{\p}{\p\ln\mu^2}\left(\ln\frac{\mu^2}{s}\right)^2,
\ee
where the factors $1/2$ are the symmetry factors for the graphs
$\g_1^{(s)}$ and $\g_2^{(s)}$,
$$
\bt_t=\frac{g^2}{2}\g_1^{(t)},\;\;
\bt_u=\frac{g^2}{2}\g_1^{(u)}
$$
are the $t$ and $u$ contributions to the beta-function.
The coefficient $c(\g_2)$ is to be found.
In our notations the derivative in $g$ is the insertion of the
corresponding graph
$$
g^2\g_1^{(t)}\left(\frac{\p}{\p g}g\right)g
\g_1^{(s)}
=g^3\g_2^{(s)}.
$$
Thus the left hand side of (\ref{bet}) is
$$
\frac{g^2}{2}(\g_1^{(t)}+\g_1^{(u)})\left(\frac{\p}{\p g}g\right)g
\g_1^{(s)}\frac{1}{2}\ln\frac{\mu^2}{s}
=\frac{g^3}{2}\g_2^{(s)}\ln\frac{\mu^2}{s}
$$
and we find that
$$
c(\g_2)=\frac{1}{2}
$$
in accordance with (\ref{tl}).

Next example is the three-loop diagram in figure 6.
\begin{figure}[h]
\begin{center}
\leavevmode
\epsfxsize 250pt
\epsffile{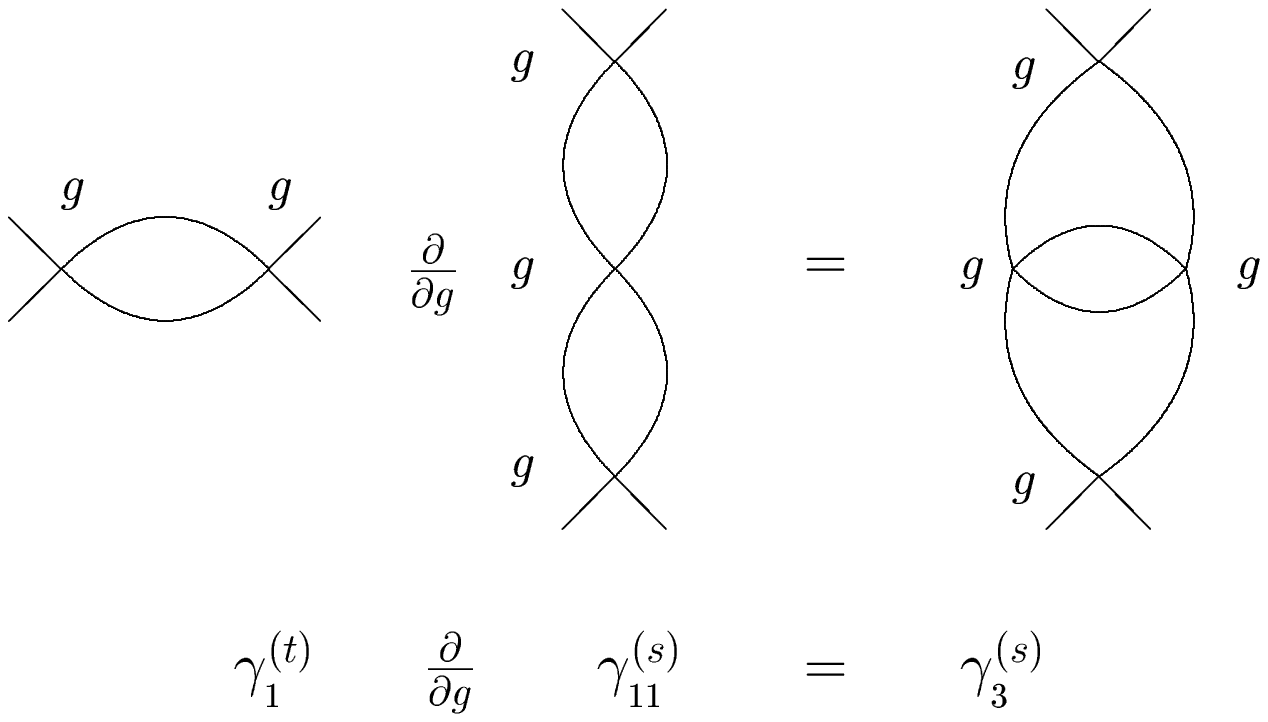}
\label{3l_ins}
\caption{\bf Insertion of  $\g_{_1}^{(t)}$ into  $\g_{_{11}}^{(s)}$}
\end{center}
\end{figure}

The corresponding equation is
\be
(\bt_t+\bt_u)\left(\frac{\p}{\p g}g\right)g^2 \g_{11}^{(s)}
\frac{1}{4}\left(\ln\frac{\mu^2}{s}\right)^2
=\frac{c(\g_3)}{4}g^4\g_3^{(s)}
\frac{\p}{\p\ln\mu^2}\left(\ln\frac{\mu^2}{s}\right)^3,
\ee
where $1/4$ is the symmetry factor for the graphs
$\g_{11}^{(s)}$ and $\g_3^{(s)}$.

Since
$$
\frac{g^2}{2}(\g_1^{(t)}+\g_1^{(u)})
\left(\frac{\p}{\p g}g\right)g^2 \g_{11}^{(s)}
=g^4\g_3^{(s)},
$$
we find that
$$
c(\g_3)=\frac{1}{3}.
$$
Again the result coincides with the previous result
(\ref{3l}).

In general, the RG invariance in the space of graphs enables one to
find the leading RG logs recursively.

Let $\g_{n+1}$ be an $(n+1)$-loop graph.
If the graph contributes to the leading logs, then it may be
obtained by insertions of the one-loop graph into some $n$-loop graphs.
In order to find these $n$-loop graphs we have to define the
operation of contraction.

Let $\g$ be a graph and let $\td{\g}\subset\g$ be a connected
subgraph. We denote by $\g/\td{\g}$ the graph $\g$ where $\td{\g}$
is contracted to a point, i.e. the graph
$\g/\td{\g}$
contains the edges and the vertices of $\g$ which do not belong to
$\td{\g}$, external edges of $\td{\g}$, and a new vertex $A$
instead of internal edges and vertices of $\td{\g}$; the (former)
external edges of $\td{\g}$ meet in this new vertex $A$.\\
For example,
\bea
\g_2/\g_1=\g_1,\\
\g_3/\g_1=\g_{11}.
\eea

Using the RG invariance in the space $\Hd$ we find
the recursive formula for the coefficients of the leading
logarithms
\be
\frac{n+1}{S(\g_{n+1})}\;\;c(\g_{n+1})
=\frac{1}{2}\;\sum_{i=s,t,u}\;\;\sum_{\g_n=\g_{n+1}/\g_1^{(i)}}\;\;
\frac{1}{S(\g_n)}\;\;c(\g_{n}),
\ee
where the factors have the following origin or definition\\
$n+1$ -- differentiation of the logarithm,\\
$S(\g)$ -- symmetry factor of the graph $\g$,\\
$c(\g)$ -- coefficient of the leading RG logarithm for $\g$,\\
$\frac{1}{2}=\frac{1}{S(\g_1)}$ -- coefficient in the one-loop beta-function,\\
$i=s,t,u$ -- orientation of the graphs,\\
$\g_n=\g_{n+1}/\g_1^{(i)}$ -- $n$-loop diagram obtained by
the contraction of the one-loop subgraph $\g_1$ in the $(n+1)$-loop graph
$\g_{n+1}$.

The last formula may be further simplified.
We note, that given $\g_{n+1}$ and $\g_n$
\be
\sum_{i:\g_1^{(i)}\ra\g_n=\g_{n+1}}
\frac{S(\g_{n+1})}{S(\g_1)S(\g_n)}=1,
\ee
where $\g_1^{(i)}\ra\g_n$ is the insertion of
$\g_1^{(i)}$ into $\g_n$ in a definite vertex
and the sum is over all possible orientations $i$
that give $\g_{n+1}$ after the
insertion.

There are two cases
\begin{enumerate}
  \item
$S(\g_1)S(\g_n)=2S(\g_{n+1})$ -- in this case we destroy one symmetry in
$\g_n$ by inserting $\g_1$, hence there are two possible orientations of
$\g_1$ that give the same $\g_{n+1}$
(these orientations are altered by the symmetry).
  \item
$S(\g_1)S(\g_n)=S(\g_{n+1})$ -- no symmetry appears or disappears, hence
there is only one orientation of $\g_1$ which gives $\g_{n+1}$.
\end{enumerate}
We stress here that, compared to the Connes and Kreimer definitions
\cite{CK1},
we fix the orientation of external edges in $\g_n$ and $\g_{n+1}$ while
inserting $\g_1$, that is why there is only one vertex in $\g_n$ where we
can insert $\g_1$ in order to obtain $\g_{n+1}$.

Now the formula takes the form
\be\lb{c2}
c(\g_{n+1})
=\frac{1}{n+1}\;\sum_{\g_n=\g_{n+1}/\g_1}\;
c(\g_{n}).
\ee
The coefficients $c(\g)$ defined
recursively by (\ref{c2}) should coincide with $c(\g)$ in formula
(\ref{c1}).
We will not prove the assertion for arbitrary graphs, but we will present
some nontrivial examples.

First of all, if there is only one $\g_1\subset\g_{n}$, then
$$
c(\g_{n})=\frac{1}{n}c(\g_{n-1}),
$$
if there is only one $\g_1\subset\g_{n-1}$ etc., then
$$
c(\g_{n})=\frac{1}{n!}
$$
and we get the right answer for the graph with only one chain of divergent
subgraphs (compare with (\ref{1ch})).

\section{Five-loop example}

Let us consider the five-loop diagram and find the coefficients
for the leading RG logarithm using the two formulas
(\ref{c1}) and (\ref{c2}).

\begin{figure}[h]
\begin{center}
\leavevmode
\epsfxsize 70pt
\epsffile{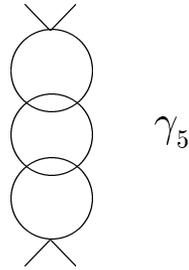}
\caption{\bf Five-loop diagram $\g_{_5}$}
\end{center}
\end{figure}

In order to use (\ref{c1}) we have to find the maximal trees for $\g_5$.

The numbers of the loops in the sub-diagrams may be organized
in the triangular (see figure 8).

\begin{figure}[h]
\begin{center}
\leavevmode
\epsfxsize 70pt
\epsffile{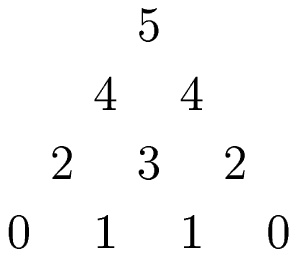}
\caption{\bf Loop numbers of the sub-diagrams in $\g_5$}
\end{center}
\end{figure}

Where '5' corresponds to the diagram and denotes the number of
loops in the diagram.
Two '4' in the second  line correspond to the two maximal
sub-diagrams which contain 4 loops.
2, 3, 2 correspond to the 2, 3, 2-loop sub-diagrams.
Note, that any of the two 2-loop sub-diagrams belongs only to one
of the 4-loop sub-diagrams, whereas the 3-loop sub-diagram belongs to
both 4-loop sub-diagrams.
The sub-diagrams which belong to the 4-loop sub-diagram
correspond to the numbers in the sub-triangular with '4' on
the top.

The maximal trees in this case correspond to the following subsets
of numbers in the triangular

\begin{figure}[h]
\begin{center}
\leavevmode
\epsfxsize 350pt
\epsffile{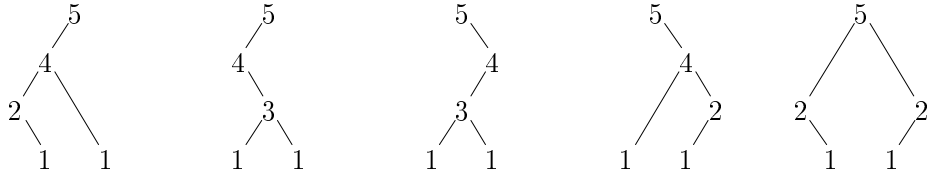}
\caption{\bf Maximal trees for $\g_5$}
\end{center}
\end{figure}

By the definition of the rooted tree the diagram (number 5)
corresponds to the root, the other numbers (sub-diagrams)
correspond to the nodes of the tree,
the links between the numbers correspond to the edges of the tree,
they represent how the diagrams belong to each other.
For example, let the left '4' correspond to the lower 4-loop
sub-diagram.
Then the left tree contains the diagram itself, the lower 4-loop
sub-diagram, the lower 2-loop sub-diagram, and the two 1-loop
sub-diagrams.
The lower 1-loop sub-diagram belongs to the 2-loop sub-diagram
and the upper 1-loop sub-diagram
belongs to the 4-loop sub-diagram.

The numbers corresponding to the trees are
$$
\frac{1}{5\cdot 4\cdot 2},\;\;
\frac{1}{5\cdot 4\cdot 3},\;\;
\frac{1}{5\cdot 4\cdot 3},\;\;
\frac{1}{5\cdot 4\cdot 2},\;\;
\frac{1}{5\cdot 2\cdot 2}.
$$
The sum
$$
c(\g_5)=\frac{2}{5\cdot 3}
$$
In the recursive formula (\ref{c2})
there are two terms which correspond to the diagrams obtained by
the contraction of one of the two 1-loop sub-diagrams.
The diagrams obtained by the contraction are shown in figure 10.

\begin{figure}[h]
\begin{center}
\leavevmode
\epsfxsize 200pt
\epsffile{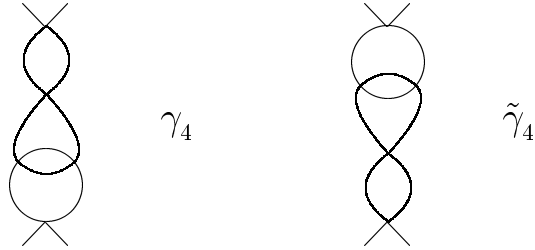}
\caption{\bf Contractions of the one-loop sub-diagrams in $\g_5$}
\end{center}
\end{figure}

These diagrams are reducible, thus
$$
c(\g_4)=c(\td{\g}_4)=c(\g_3)c(\g_1)=\frac{1}{3},
$$
here $\g_3$ is the 3-loop diagram (see figure 4)
and $\g_1$ is the 1-loop diagram.
The recursive formula in this case
$$
c(\g_5)
=\frac{1}{5}\left(c(\g_4)+c(\td{\g}_4)\right)
=\frac{1}{5}\left(\frac{1}{3}+\frac{1}{3}\right)
=\frac{2}{5\cdot 3}.
$$

\begin{figure}[h]
\begin{center}
\leavevmode
\epsfxsize 70pt
\epsffile{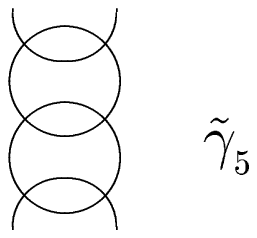}
\caption{\bf Five-loop diagram $\td{\g}_{_5}$}
\end{center}
\end{figure}

An interesting fact is that for the five-loop diagram
$\td{\g}_{_5}$, shown in figure 11,
the coefficient is the same
$$
c(\td{\g}_5)=\frac{2}{5\cdot 3}.
$$

\begin{figure}[h]
\begin{center}
\leavevmode
\epsfxsize 200pt
\epsffile{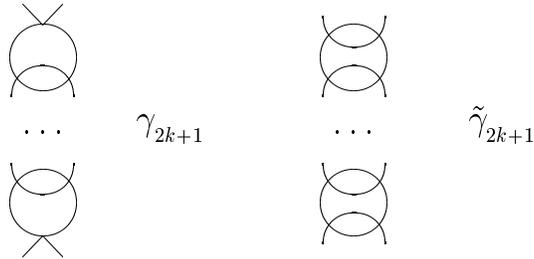}
\caption{\bf (2k+1)-loop diagrams that have the same leading logs}
\end{center}
\end{figure}

This property seems to be true for any diagrams
of the form shown in figure 12.

For example,
\bea
c(\g_3)&=&c(\td{\g}_3)=\frac{1}{3},\\
c(\g_5)&=&c(\td{\g}_5)=\frac{2}{5\cdot 3},\\
c(\g_7)&=&c(\td{\g}_7)=\frac{17}{7\cdot 5\cdot 3\cdot 3}.
\eea
This is an interesting unexpected
result that the leading logs for the diagrams in figure
12 coincide.

\section{Conclusion and acknowledgements}

We propose a decomposition of singular integrals which is used
to find the leading RG logs in $\vp^4$ theory.
We have considered only the leading RG logs,
but we have given a general
prescription of extracting RG and non-RG parts,
thus the idea may be used to study the
next-to-leading RG logs as well as the non-RG
logs.
At the moment the beta-function in $\vp^4$ theory is known up to five loops
\cite{Gorishnii},
thus it should be possible to find four next-to-leading RG logs.

In the last parts we obtained the same result for the leading RG logs using
the Lie algebra of graphs.
This approach also has a straightforward generalization for the
next-to-leading logs.

The author is thankful to K.Selivanov and V.Belokurov for constant attention
to the work,
and to D.Kazakov, V.Rubakov, and I.Ginzburg for inspiring discussions.

The work is supported in part by RFBR grant 03-01-00371
and INTAS grant 00-334.


\newpage

 \begin{thebibliography}{40}

\bibitem{Tkachev}
A.~N.~Kuznetsov, F.~V.~Tkachov and V.~V.~Vlasov,
Techniques Of Distributions In Perturbative Quantum Field Theory.
1. Euclidean Asymptotic Operation For Products Of Singular Functions,
hep-th/9612037.


\bibitem{Grozin}
A.~G.~Grozin,
Progress in multiloop calculations,
Nucl.\ Instrum.\ Meth.\ A {\bf 502}, 815 (2003),
hep-ph/0211351.



 \bibitem{Bog}
N.N.Bogoliubov, D.V.Shirkov,
Introduction to the Theory of Quantized Fields,
3rd ed., Wiley-Interscience, 1980.

 \bibitem{Itsyk}
C.Itsykson, J.B.Zuber, Quantum Field Theory,
Mc Grow-Hill, 1980.

 \bibitem{Col}
J.Collins, Renormalization, Cambridge Univercity Press, Cambridge,
1984.




\bibitem{Delam}
B.~Delamotte,
A hint of renormalization,
hep-th/0212049.


 \bibitem{CK1}
 A.Connes, D.Kreimer,
 Renormalization in quantum field theory and
 the Riemann-Hilbert problem I: the Hopf algebra structure of
 graphs and the main theorem, hep-th/9912092.


 \bibitem{CK2}
 A.Connes, D.Kreimer,
 Renormalization in quantum field theory and the Riemann-Hilbert problem II:
 the $\beta$-function, diffeomorphisms and the renormalization group,
 hep-th/0003188.

\bibitem{'tHdimreg}
G.~'t Hooft,
Dimensional Regularization And The Renormalization Group,
Nucl.\ Phys.\ B {\bf 61}, 455 (1973).

\bibitem{Gelfand}
I.M.Gelfand, G.E.Shilov,
Generalized Functions,
vol. 1, Academic: New York, 1964.

\bibitem{Zav}
O.I.Zav'ialov, Renormalized Feynman Diagrams, Moscow, Nauka, 1979
(in Russian).

\bibitem{Weinberg}
S.~Weinberg,
High-Energy Behavior In Quantum Field Theory,
Phys.\ Rev.\  {\bf 118}, 838 (1960).



\bibitem{Ginzburg}
I.~F.~Ginzburg, A.~V.~Yefremov and V.~G.~Serbo,
Asymptotic Of Plane Graphs In Spinor Theory,
Yad.\ Fiz.\  {\bf 9}, 451 (1969).




\bibitem{Zim:forest}
W.~Zimmermann,
Commun.\ Math.\ Phys.\  {\bf 15}, 208 (1969)
[Lect.\ Notes Phys.\  {\bf 558}, 217 (2000)].


\bibitem{Peskin}
M.E.Peskin, D.V.Schroeder, An introduction to Quantum Field
Theory, Addison-Wesley Publishing Company, 1995.

 \bibitem{GMS}
 A.Gerasimov, A.Morozov, K.Selivanov,
 Bogolubov's Recursion and Integrability of Effective Actions,
 Int.J.Mod.Phys.A16:1531-1558,2001,
 hep-th/0005053


\bibitem{Connes:2002ui}
A.~Connes and D.~Kreimer,
Insertion and elimination: The doubly infinite Lie algebra of Feynman  graphs,
Annales Henri Poincare {\bf 3}, 411 (2002),
hep-th/0201157.

\bibitem{Kreimer:2002qy}
D.~Kreimer,
New mathematical structures in renormalizable quantum field theories,
Annals Phys.\  {\bf 303}, 179 (2003)
[Erratum-ibid.\  {\bf 305}, 79 (2003)],
hep-th/0211136.


\bibitem{Kr:overlap}
D.~Kreimer,
On overlapping divergences,
Commun.\ Math.\ Phys.\  {\bf 204}, 669 (1999),
hep-th/9810022.


\bibitem{Connes:tree}
A.~Connes and D.~Kreimer,
Hopf algebras, renormalization and noncommutative geometry,
Commun.\ Math.\ Phys.\  {\bf 199}, 203 (1998),
hep-th/9808042.



\bibitem{Gorishnii}
S.~G.~Gorishnii, S.~A.~Larin, F.~V.~Tkachov and K.~G.~Chetyrkin,
Five Loop Renormalization Group Calculations In The G Phi**4 In Four-Dimensions Theory,
Phys.\ Lett.\ B {\bf 132}, 351 (1983).


 \end{thebibliography}
\end{document}